\documentclass[10pt,conference]{IEEEtran}

\usepackage{cite}
\usepackage{amsmath}
\usepackage{amsfonts}
\usepackage{amssymb}
\usepackage{mathrsfs}
\usepackage[pdftex]{graphicx}
\usepackage{color, colortbl}
\usepackage[dvipsnames]{xcolor}
\usepackage{pgfplots}
\usepackage{tikz}
\usepackage{tikzscale}
\usepackage{enumitem}

\newcommand{\subparagraph}{}
\usepackage{titlesec}
\usepackage{fancyhdr}

\AtBeginEnvironment{tabular}{\scriptsize\vspace{-3mm}}
\AtBeginEnvironment{table}{\vspace{-3mm}}
\AtEndEnvironment{table}{\vspace{-3mm}}

\AtBeginEnvironment{figure}{\vspace{-3mm}}
\AtEndEnvironment{figure}{\vspace{-3mm}}

\titlespacing*{\section}
{0pt}{3mm}{1mm}
\titlespacing*{\subsection}
{0pt}{2mm}{1mm}

\newcommand{\etal}{\textit{et al.} }
\newcommand{\cf}{\textit{cf.}, }
\newcommand{\eg}{\textit{e.g.}, }
\newcommand{\ie}{\textit{i.e.}, }
\newcommand{\etc}{\textit{etc}.}

\newcommand{\fig}[1]{Fig.~\ref{#1}}
\newcommand{\tab}[1]{Table~\ref{#1}}

\newcommand{\red}[1]{\textcolor{red}{#1}}

\usetikzlibrary{arrows.meta, fit, backgrounds, shapes.geometric}
\tikzset{%
  >={Latex[width=2mm,length=2mm]},
         base/.style = {rectangle, rounded corners, draw=black,
                           minimum width=3cm, minimum height=1cm,
                           text centered, font=\rmfamily},
         terminal/.style = {base, circle, minimum size=1.5cm,font=\rmfamily}
}

\pgfplotsset{compat=1.14}

\IEEEoverridecommandlockouts

\begin{document}
\title{A Graph-Based Machine Learning Approach for Bot Detection}

\author{
	\IEEEauthorblockN{
		Abbas Abou Daya, Mohammad A. Salahuddin, Noura Limam, and Raouf Boutaba\\
	}
	\IEEEauthorblockA{
	David R. Cheriton School of Computer Science, University of Waterloo, Ontario, Canada\\
	\texttt{\{aaboudaya, mohammad.salahuddin, noura.limam, rboutaba\}@uwaterloo.ca}}
}

\maketitle
\thispagestyle{fancy}

\begin{abstract}
 Bot detection using machine learning (ML), with network flow-level features, has been extensively studied in the literature. However, existing flow-based approaches typically incur a high computational overhead and do not completely capture the network communication patterns, which can expose additional aspects of malicious hosts. Recently, bot detection systems which leverage communication graph analysis using ML have gained attention to overcome these limitations. A graph-based approach is rather intuitive, as graphs are true representations of network communications. In this paper, we propose a two-phased, graph-based bot detection system which leverages both unsupervised and supervised ML. The first phase prunes presumable benign hosts, while the second phase achieves bot detection with high precision. Our system detects multiple types of bots and is robust to zero-day attacks. It also accommodates different network topologies and is suitable for large-scale data.

\end{abstract}
\section{Introduction}\label{sec:intro}


Undoubtedly, organizations are constantly under security threats, which not only cost billions of dollars in damage and recovery, but also detrimentally affect their reputation. A botnet-assisted attack is a widely known threat to these organizations. According to the U.S. Federal Bureau of Investigation, ``Botnets caused over \$9 billion in losses to U.S. victims and over \$110 billion globally.'' 
 The most infamous attack, Rustock, infected 1 million machines, sending up to 30 billion spam emails a day~\cite{caballero2011measuring}. More recently, 
WannaCry 
resulted in 
data breach from over 230,000 computers in 150 countries~\cite{ehrenfeld2017wannacry}. 
Undeniably, in the face of a cyber arms race, attackers constantly find clever ways to sabotage networks using botnets, most importantly via zero-day attacks~\cite{boutaba2018comprehensive}. 

A botnet is a collection of \textit{bots}, agents in compromised hosts, controlled by botmasters via command and control (C2) channels. A malevolent adversary controls the bots through 
botmaster, which could be distributed across several agents that reside within or outside the network. Hence, bots can be used for tasks ranging from distributed denial-of-service (DDoS), to massive-scale spamming, to fraud and identity theft. 
While bots thrive for different sinister purposes, they exhibit a similar behavioral pattern when studied up-close. The intrusion kill-chain~\cite{hutchins2011intelligence} dictates the general phases a malicious agent goes through in-order to reach and infest its target. 

Detection of bots can be largely achieved via 
intrusion detection systems (IDSs), 
which can be broadly classified into \textit{signature-based} and \textit{anomaly-based}~\cite{khattak2014taxonomy}. Signature-based methods use \textit{pre-computed} hashes of existing malware binaries. They scale well and efficiently detect known threats. 
However, 
they require frequent database updates and can be subverted by unknown or modified attacks, such as zero-day attacks and polymorphism \cite{khattak2014taxonomy,debar1999towards}. This undermines their suitability for bot detection.
Anomaly-based methods overcome these limitations~\cite{boutaba2018comprehensive,creech2014semantic}. They establish a baseline of normal behavior for the protected system, and model a decision engine that alerts on any divergence or statistical deviations from the norm. \textit{Machine learning} (ML) is an ideal technique to automatically capture the normal behavior of a system. Its use has boosted the scalability and accuracy of IDSs~\cite{boutaba2018comprehensive,creech2014semantic}. 

An important step prior to learning, or training a ML model, is feature extraction. These features act as discriminators for learning and inference, reduce data dimensionality, and increase the accuracy of ML models. The most commonly employed features in bot detection are flow-based (\eg source and destination IPs, protocol, number of packets sent and/or received, etc.). However, these features do not completely capture the communication patterns that can expose additional aspects of malicious hosts. In addition, flow-level models can incur a high computational overhead, and can also be evaded by tweaking behavioral characteristics \eg by changing packet structure~\cite{venkatesh2015botspot}. 

\textit{Graph-based} features, derived from flow-level information to reflect the true behaviour of hosts, are an alternate that overcome these limitations. We show that incorporating graph-based features into ML yields robustness against complex communication patterns and unknown attacks. Moreover, it allows for cross-network ML model training and inference.  
The major contributions of this paper are as follows:
\begin{itemize}[leftmargin=*]
\vspace{-1mm}
\item We propose an anomaly-based approach for bot detection that is protocol agnostic, robust to zero-day attacks, and suitable for large datasets. 
\item We show the limitations of stand-alone supervised learning. 
Therefore, we employ a two-phased ML approach that leverages both supervised and unsupervised learning. The first phase filters presumable benign hosts. 
This is followed by the second phase 
on the pruned hosts, to achieve bot detection with high precision.
\item We use graph-based features and evaluate various ML techniques. The graph-based features, derived from network flows, overcome severe topological effects that can skew bot behavior, exacerbating ML prediction. Furthermore, these features allow to combine data from different networks and promote spatial stability~\cite{jin2012modular} in the models.
\end{itemize}
The rest of the paper is organized as follows. In Section~\ref{sec:background}, we present a background on bot detection and highlight limitations of the state-of-the-art. Our system design is delineated in Section~\ref{sec:design}. We discuss the results of our experiments in Section~\ref{sec:experiments}. In Section~\ref{sec:conclusion}, we conclude with a summary of our contributions and instigate future research directions.
\section{Related Works}\label{sec:background}


Bot(net) detection has been an active area of research and has generated a substantial body of work. Most existing bot detection techniques employ methods for detecting C2 channels based on the statistical features of packets and flows~\cite{binkley2006algorithm, karasaridis2007wide, goebel2007rishi, strayer2008botnet, villamarin2008identifying, gu2008botminer, lu2009botcop, zeidanloo2010botnet, saad2011detecting, zhang2011detecting, lu2011clustering, choi2012identifying,  zhao2013botnet}. Solutions like \cite{binkley2006algorithm, karasaridis2007wide} are focused on specific communication protocols, such as IRC, providing narrow-scoped solutions. On the other hand, Botminer \cite{gu2008botminer} is a protocol-independent solution, which assumes that bots within the same botnet are characterized by similar malicious activities and communication patterns. This assumption is an over simplification, since botnets often randomize topologies \cite{khattak2014taxonomy} and communication patterns as we observe in newer malware, such as Mirai~\cite{antonakakis2017understanding}. Therefore, it is evident that a non-protocol-specific, less evadable detection method is desired. 

Furthermore,~\cite{saad2011detecting,zhao2013botnet} have exploited ML-driven anomaly detection with traffic-based statistical features, for detecting known and unknown attacks with low error rates. However, such techniques require that all flows are compared against each other to detect C2 traffic, which incurs a high computational overhead. In addition, they are unreliable, as they can be evaded with encryption and by tweaking flow characteristics~\cite{venkatesh2015botspot}. 
Graph-based approaches, where graphs are extracted from network flows and host-to-host communication patterns, overcome these limitations~\cite{collins2007hit, nagaraja2010botgrep, francois2011botcloud, hang2013entelecheia, venkatesh2015botspot, henderson2012rolx, ding2012intrusion, wang2014botnet, franccois2011bottrack, jaikumar2015graph}. Other approaches~\cite{zhuang2017peerhunter, zhuang2018enhanced} rely on rule-based host classification and botnet detection methods, where pre-defined thresholds are used to discriminate between benign and suspicious hosts. 
They are static and prone to evasion. 

On the other hand, outlier- and anomaly-based methods are generally more robust.
BotGM~\cite{lagraa2017botgm} uses a statistical technique, the inter-quartile method, for outlier detection. Their results exhibit moderate accuracy with low FPs based on different windowing parameters. However, it generates multiple graphs from a selection of network flows. For every pair of unique IPs, a graph is constructed, such that every node in the graph represents a unique 2-tuple of source and destination ports, with edges signifying the time sequence of communication. This entails high overhead and will not scale for large datasets.

Chowdhury \etal \cite{chowdhury2017botnet} use ML for clustering the nodes in a graph, with a focus on dimensionality and topological characterization. Their assumption is that most benign hosts will be grouped in the same cluster due to similar connection patterns, hence can be eliminated from further analysis. Such a crucial reduction in nodes effectively minimize detection overhead. However, their graph-based features are plagued by severe topological effects (\cf Section~\ref{sec:experiments}). They use statistical means and user-centric expert opinion to tag the remaining clusters as malicious or benign. Nevertheless, leveraging expert opinion can be cumbersome, error prone and infeasible for large datasets.

Graph-based approaches using ML for bot detection are intuitive and show promising results. In this paper, we propose an anomaly-, graph-based bot detection system, which is protocol agnostic \ie it detects bots regardless of the protocol. We employ graph-based features in a two-phased ML approach, which is robust to zero-day attacks, spatially stable, and suitable for large datasets. 
\section{System Design}\label{sec:design}
\begin{figure}[htb]
  \centering
  \resizebox{0.35\textwidth}{!}{%
  \begin{tikzpicture}[node distance=4cm, every node/.style={fill=white, font=\rmfamily}, align=center]
    \node (flow_ingestion) [base, anchor=north west] {Flow \\ Ingestion};
    \node (graph_transform) [base, right of =flow_ingestion] {Graph \\ Transform};
    \node (feature_extraction) [base, below of=graph_transform, yshift=2cm] {Feature \\ Extraction};
    \node (fnorm) [base, left of=feature_extraction] {Feature \\ Normalization};
    \node (phase1) [base, below of=fnorm, yshift=1cm] {Phase 1 \\ (Unsupervised)};
    \node (phase2) [base, right of=phase1] {Phase 2 \\ (Supervised)};
    \node (model) [base, fill=red!40, below of=phase1, xshift=2cm] {Inference};
    \node (host) [terminal, left of=model] {Host};
    \node (benign) [terminal, right of=model, yshift=1cm] {Benign};    
    \node (bot) [terminal, right of=model, yshift=-1cm] {Bot};
    \draw[->] (flow_ingestion) -- (graph_transform);
    \draw[->] (graph_transform) -- (feature_extraction);
    \draw[->] (feature_extraction) -- (fnorm);
    \draw[->] (fnorm) -- (phase1);
    \draw[->] (phase1) -- (phase2);
    \draw[->] (phase1) -- ++(0,-2) -| (model);
    \draw[->] (phase2) -- ++(0,-2) -| (model);
    \draw[->] (host) -- (model);
    \draw[->] (model) -- ++(2,0) |- (benign);
    \draw[->] (model) -- ++(2,0) |- (bot);
    
    \begin{scope}[on background layer]
      \node[draw=blue, fill=blue!40, inner sep=2mm, rounded corners, label=below:Data Bootstrap,fit=(flow_ingestion)(fnorm) (feature_extraction) (fnorm)] {};
  	\end{scope}
 
    \begin{scope}[on background layer]
      \node[draw=orange, fill=orange!40, inner sep=2mm, rounded corners, label=below:Model Training,fit=(phase1)(phase2)] {};
  	\end{scope}
  \end{tikzpicture}}
  \caption{Components of the bot detection system}
  \label{fig:design}
\end{figure}
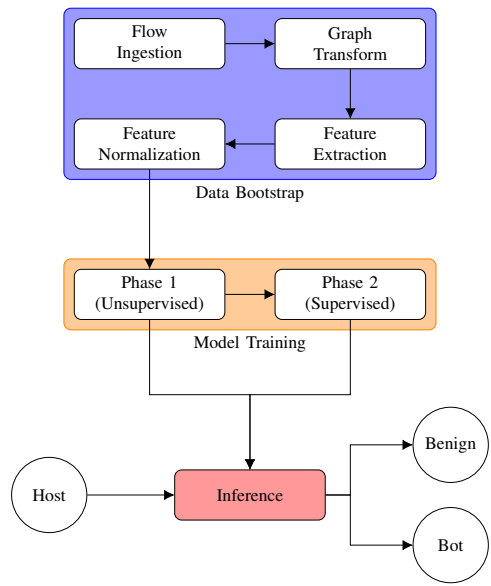

Our bot detection system consists of 3 major components, as depicted in~\fig{fig:design}. These components pertain to data preparation and feature extraction, model training, and inference.  
\subsection{Dataset bootstrap}
\subsubsection{Flow ingestion}
The input to the system are bidirectional network flows. These flows are transformed into a set $T$ that contains 4-tuple flows $t_i=\{sip_i,~srcpkts_i,~dip_i,~dstpkts_i\}$. Where $sip_i$ is the source IP address that uniquely identifies a source host, $srcpkts_i$ quantifies the number of data packets sent by $sip_i$ to $dip_i$, the destination host IP address. The number of destination packets, $dstpkts_i$, is the number of data packets sent by $dip_i$ to $sip_i$.

Set $A$ is a set of tuples that have exclusive source and destination hosts. That is, if multiple tuples have the same source and destination hosts, they are reduced to form an aggregated exclusive tuple $a_x \in A$, such that $a_x=\{sip_x,~srcpkts_x,~dip_x,~dstpkts_x\}$. Therefore, if two tuples $t_i, t_j \in T$ have the same source and destination hosts \ie $sip_x = sip_i = sip_j$ and $dip_x = dip_i = dip_j$, then number of source and destination packets are aggregated in $a_x$, such that
\begin{equation}
srcpkts_x = \sum_{t_k\in T~|~sip_x = sip_k, dip_x = dip_k} srcpkts_k
\end{equation}
\begin{equation}
dstpkts_x = \sum_{t_k \in T~|~sip_x = sip_k, dip_x = dip_k} dstpkts_k.
\end{equation}


\subsubsection{Graph Transform}
The system creates a graph $G(V,~E)$, where $V$ is a set of nodes and $E$ is a set of directed edges $e_{i,j}$ from node $v_i$ to node $v_j$ with weight $|e_{i,j}|$. The set of nodes $V$ is a union of hosts from set $A$, such that

\begin{equation} 
\label{graph_verts} 
V = \bigcup_{\forall a_x \in A}\{sip_x \cup dip_x\}. 
\end{equation}
For every $a_x \in A$, there exist directed edges $e_{i,j}$ and $e_{j,i}$ from $v_i$ to $v_j$ and $v_j$ to $v_i$, respectively, such that $sip_x = v_i$ and $dip_x = v_j$. Therefore,

\begin{equation} 
\label{graph_edges} 
E = \bigcup_{\forall a_x \in A}\{(sip_x,~dip_x) \cup (dip_x,~sip_x)\}. 
\end{equation}
The weights $|e_{i,j}|$ and $|e_{j,i}|$ of edges $e_{i,j}$ and $e_{j,i}$ are $srcpkts_x$ and $dstpkts_x$, respectively. Moreover, if there exists a reverse tuple $a_y \in A~|~dip_y = v_i,~sip_y = v_j$, then $|e_{i,j}| = srcpkts_x~+~dstpkts_y$ and $|e_{j,i}| = dstpkts_x~+~srcpkts_y$.

\vspace{5pt}
\subsubsection{Feature Extraction}
The system creates the required \textit{graph-based} feature set for the ML model. Features are intrinsic to the success of the model that should genuinely represent and discriminate host behavior, especially bot behavior. We study the following set of commonly used graph-based features. 
\begin{itemize}[leftmargin=*]
\item \textbf{\textit{In-Degree (ID)}} and \textbf{\textit{Out-Degree (OD)}}---The in-degree, $f_{i,0}$, and out-degree, $f_{i,1}$, of a node $v_i \in V$ are the number of its ingress and egress edges, respectively. 
\begin{equation} 
\label{feat_count} 
\mathcal{F}(e_{i,j}) = \begin{cases}
    1,& \text{if } e_{i,j} \in E\\
    0,& \text{otherwise}
\end{cases}
\end{equation}
\begin{equation}
\label{feat_id} 
f_{i,0} = \sum_{v_j \in V,~v_i \neq v_j} \mathcal{F}(e_{j,i})~~~~~\forall v_i \in V
\end{equation}
\begin{equation}
\label{feat_od} 
f_{i,1} = \sum_{v_j \in V,~v_i \neq v_j} \mathcal{F}(e_{i,j})~~~~~\forall v_i \in V
\end{equation}
These features play an important role in the behavior of a host. Though, a higher ID for a host makes it a point of interest, often nodes with high ID offer commonly demanded service. Therefore, observing ID alone may not signify malicious activity. For example, a gateway is a central point of network communication, but it is not necessarily a malicious endpoint. 
Intuitively, bots attempting to infect the network will tend to have higher ID than benign hosts. Similarly, OD is also an intrinsic feature. Typically, in the reconnaissance stage of the intrusion kill-chain, bots attempt to mass-survey the network, which can be captured via OD.


\item \textit{\textbf{In-Degree Weight (IDW)}} and \textbf{\textit{Out-Degree Weight (ODW)}}---These features augment the ID and OD of the nodes in the graph. The in-degree weight, $f_{i,2}$, and the out-degree weight, $f_{i,3}$, of a node $v_i \in V$ is the sum of all the weights of its incoming and outgoing edges, respectively. 
\begin{equation}
\label{feat_idw} 
f_{i,2} = \sum_{v_j \in V,~v_i \neq v_j,~e_{j,i} \in E} |e_{j,i}|~~~~~\forall v_i \in V
\end{equation}

\begin{equation}
\label{feat_odw} 
f_{i,3} = \sum_{v_j \in V,~v_i \neq v_j,~e_{i,j} \in E} |e_{i,j}|~~~~~\forall v_i \in V
\end{equation}
For a fine-grained differentiation, it is important to expose features that will eventually bring bots closer to each other, and discriminate bots from hosts.
IDW and ODW features add another layer of information, further alienating the malicious hosts from the benign. Similar to ID, mass-data leeching bots will tend to expose a high IDW in the action phase of the intrusion kill-chain. 
Similarly, the ODW is the aggregate data packets a node has sent, which can potentially expose bots that mass-send payloads to hosts in a network.






\item \textbf{\textit{Betweenness Centrality (BC)}}---The betweenness centrality of a node $v_i \in V$, inspired from social network analysis, is a measure of the number of shortest paths that pass through it, such that

\vspace{-5mm}
\begin{equation} 
\label{feat_bc} 
f_{i,4} = \sum_{v_j, v_k \in V,~v_i \neq v_j \neq v_k} \frac{\sigma_{v_j v_k}(v_i)}{\sigma_{v_j v_k}}~~~~~\forall v_i \in V.
\end{equation}

Where $\sigma_{v_j v_k}$ is the total number of shortest paths between node pairs $v_j,~v_k \in V$, and $\sigma_{v_j v_k}(v_i)$ is the number of shortest paths that pass through $v_i$. This feature has a high computational overhead with $O(|V|.|E|+|V|^2.\log |V|)$ time complexity~\cite{brandes2001faster}. However, it can alienate bots early on as they attempt their first connections. This is when the bots exhibit low IDW and ODW. Thus, it would be more favorable for the shortest paths in the network to pass through the host. Likewise, when the IDW and ODW increase, the BC of a node decreases immensely, as it is less favored for being included in shortest paths.

\item \textbf{\textit{Local Clustering Coefficient (LCC)}}---Unlike BC, local clustering coefficient has a lower computational overhead, and it quantifies the neighborhood connectivity of a node $v_i \in V$, such that
\begin{equation} 
\label{feat_lcc} 
f_{i,5} = \frac{\sum_{v_j, v_k \in N_i,~v_i \neq v_j \neq v_k} {\mathcal{F}(e_{j,k})}}{|N_i|(|N_i| - 1)}~~~~~\forall v_i \in V
\end{equation}
Where $N_i$ is neighborhood set for $v_i$, $\forall v_j \in N_i~|~e_{i,j} \in E \vee e_{j,i} \in E$. 
LCC feature can play an important role in discriminating malicious host's behavior. Successfully infected hosts tend to exhibit a higher LCC, as bots often deploy collaborative P2P techniques, making its adjacent host pairs strongly connected.

\item \textit{\textbf{Alpha Centrality (AC)}}---Alpha centrality, also inspired from social network analysis, is a feature that generalizes the centrality of a node $v_i \in V$. AC extends the Eigenvector centrality (EC), with the addition that nodes are also influenced by external sources. These external sources can be user-defined, according to their graphical analysis technique. 
In EC, each $v_i$ is assigned an influence score $x_i$, that is iteratively exchanged with 
adjacent nodes. In essence, EC is the relative weight of a node in the graph, such that connections to high-scoring nodes contribute more to the score of $v_i$. 
Hence, AC is given as
\begin{equation} 
\label{feat_ac} 
f_{i,6} = \alpha A^T_{i}x_{i} + e_i~~~~~\forall v_i \in V.
\end{equation}
Where $A_{i}$ is the adjacency matrix, $e_{i}$ is the external influence of node $v_i$, and $\alpha$ is influence factor that controls focus between external sources to internal influence.
AC is important for intermediate and terminal phases of the intrusion kill-chain. Early on, it may be negligible. However, as time progresses and bots perform more actions in the network, their AC will gradually increase, making it discriminative.
\end{itemize}

\subsubsection{Feature Normalization (F-Norm)}

Topological alterations can severely affect the host's behavior and pattern of communication in the graph. For example, in~\fig{fig:top_w_gw}, $g$ acts as a gateway for host $h2$ to communicate with the rest of the network (\ie hosts $h3$, $h4$ and $h5$). In this configuration, $h_2$ carries an ID of 2. In contrast,~\fig{fig:top_wo_gw} shows the topology without a gateway, where $h_2$ communicates with other hosts in the network individually. This boosts the ID of host $h_2$ to 4. To alleviate this adverse effect of different network topologies, we normalize the above \emph{base} features to incorporate neighborhood relativity.

\begin{figure}[htb]
  \centering
  \resizebox{0.32\textwidth}{!}{%
  \begin{tikzpicture}[node distance=2cm, every node/.style={fill=white, font=\rmfamily}, align=center, node/.style = {draw=black, text width=12pt, circle, minimum size=1cm,font=\sffamily}]
    \node (h1) [node] {$h_1$};v
    \node (h2) [node, right of=h1] {$h_2$};
    \node (g) [node, fill=lightgray, right of=h2] {$g$};
    \node (h3) [node, right of=g] {$h_3$};
    \node (h4) [node, above of=h3] {$h_4$};
    \node (h5) [node, right of=h4] {$h_5$};    
    \draw[<->] (h1) -- (h2);
    \draw[<->] (h2) --  (g);
    \draw[<->] (g) --  (h3);
    \draw[<->] (g) --  (h4);
    \draw[<->] (g) --  (h5);
    \draw[<->] (h3) -- (h4);
    \draw[<->] (h3) -- (h5);
    \draw[<->] (h4) -- (h5);
  \end{tikzpicture}}
  \caption{Example topology of benign hosts with a gateway}
  \label{fig:top_w_gw}
\end{figure}
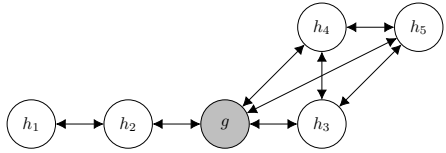
\begin{figure}[htb]
  \centering
  \resizebox{0.25\textwidth}{!}{%
  \begin{tikzpicture}[node distance=2cm, every node/.style={fill=white, font=\rmfamily}, align=center, node/.style = {draw=black, text width=12pt, circle, minimum size=1cm,font=\sffamily}]
    \node (h1) [node] {$h_1$};
    \node (h2) [node, right of=h1] {$h_2$};
    \node (h3) [node, right of=h2] {$h_3$};
    \node (h4) [node, above of=h3] {$h_4$};
    \node (h5) [node, right of=h4] {$h_5$};    
    \draw[<->] (h1) -- (h2);
    \draw[<->][<->] (h2) -- (h3);
    \draw[<->] (h2) -- (h4);
    \draw[<->] (h2) -- (h5);
    \draw[<->] (h3) -- (h4);
    \draw[<->] (h3) -- (h5);
    \draw[<->] (h4) -- (h5);
  \end{tikzpicture}}
  \caption{Example topology of benign hosts without a gateway}
  \label{fig:top_wo_gw}
\end{figure}
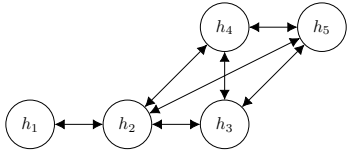

To control the overhead of computing these \textit{normalized} features, the neighborhood set $N_i$ for $v_i \in V$ is restricted to a depth $D$. The mean of $j$ features  for $v_i$ across its neighbors $v_k \in N_i$ are computed. Each feature for $v_i$ is then normalized by incorporating neighborhood relativity. Thus, features relative to their neighborhood mean is given as
\begin{equation} 
\label{feat_nfr_mean} 
\mu_{i,m} = \frac{\sum_{v_k \in N_i}f_{k,m}}{|N_i|}~~~~~\forall v_i \in V,~0 \leq m \leq j
\end{equation}

\begin{equation} 
\label{feat_nfr} 
f_{i,m} = \frac{f_{i, m}}{\mu_{i, m}}~~~~~\forall v_i \in V,~0 \leq m \leq j.
\end{equation}

After normalizing the features (with $D=2$) 
for $h_2$ and $h_4$ with gateway, their IDs change from $2$ to $0.8$ and $3$ to $1.1$, respectively. Without the gateway, their IDs change from $4$ to $1.6$ and $3$ to $1.1$, respectively. As aforementioned, normalization attempts to make hosts of the same nature look similar, 
making the topological alterations less severe. Similarly, in situations where network data is recorded over varying time intervals, IDW and ODW tend to increase substantially with larger time intervals. By normalizing features, the effect of time also diminishes.

\subsection{Model Training}

The model is trained to accept graph-based features as input and learn to distinguish between malicious and benign hosts. Two learning phases are involved as explained below.
\subsubsection{Phase 1} The first ML phase performs unsupervised learning (UL) to cluster the hosts. Generally, benign hosts exhibit \textit{similar} behavior that can be gauged by the graph-based features. 
These hosts exhibit resembling patterns in data, such as sending (OD/ODW) and receiving (ID/IDW) similar number of packets~\cite{chowdhury2017botnet}.  
Since BC, LCC and AC are directly affected by these traits, their influence can be similar for all benign hosts. This may maximize the size of the benign cluster.

This phase not only acts as a first filter for new hosts, but also significantly reduces the training data for the second phase. If a host is clustered into the benign cluster, then it is strictly benign. However, it is important to note that a malicious host can also be incorrectly clustered into a benign cluster, adversely affecting system performance. Although the objective is to maximize the size of the benign cluster, it is essential to \textit{jointly} minimize the number of bots that are co-located in this cluster. 
Various UL techniques can be deployed in this phase, including \textit{k}-Means, Density-Based Spatial Clustering (DBScan) and Self-Organizing Map (SOM). However, the classifier with the best assignment must be selected, according to the objective outlined in this phase.

\begin{figure}[htb]
  \centering
  \resizebox{0.23\textwidth}{!}{%
  \begin{tikzpicture}[node distance=2.2cm, every node/.style={fill=white, font=\sffamily \footnotesize}, align=center]
    \node (input) {$[f_0,f_1,f_2,...,f_j]_i$};
    \node (phase1) [base, below of=input, yshift=0.5cm] {Phase 1};
    \node (isbenign) [base, diamond, minimum size=1.2cm, below of=phase1] {HOB?};
    \node (phase2) [base, below of=isbenign] {Phase 2};
    \node (bot) [terminal, left of=phase2, yshift=-2cm] {Bot};
    \node (benign) [terminal, right of=phase2, yshift=-2cm] {Benign};    
    \draw[->] (input) -- (phase1);
    \draw[->] (phase1) -- (isbenign);
    \draw[->] (isbenign) -- node [yshift=0.1cm] {Yes} (phase2);
    \draw[->] (isbenign) -| node {No} (benign);
    \draw[->] (phase2) -| (benign);
    \draw[->] (phase2) -| (bot);
  \end{tikzpicture}}
  \caption{Flowchart of node classification with $i$ nodes and $j$ features}
  \label{fig:flowchart}
\end{figure}
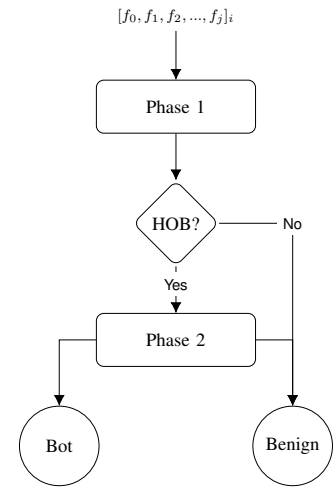

\subsubsection{Phase 2}
Phase 1 separates the dataset between hosts that are inside and outside the benign cluster. All the hosts, ideally a small number, that reside outside the benign cluster are an input to Phase 2 for further classification. Optimally, all bots should be outside the benign cluster, regardless of whether or not they are co-located in the same cluster. Depending on the number of hosts outside the benign cluster, supervised learning (SL) classifiers in this phase will exhibit varying results. 

The primary objective in this phase is to maximize recall \ie 
how many bots do not go unnoticed. 
It is proportional to the number of true positives (TPs) and inversely proportional to FNs. Various SL classifiers can be deployed to achieve this objective, such as logistic regression (LR), support vector machine (SVM), feed-forward neural network (FNN) and decision tree (DT). 
The objective from Phase 1 \ie minimize hosts outside the benign cluster (HOB), while maximizing bots outside the benign cluster (BOB), results in a minimal training dataset for Phase 2. Also, the resultant training dataset from Phase 1 is expected to be unbalanced, with a bias towards benign hosts. This may prove problematic for LR, SVM and FNN in achieving high recall rates.

\subsection{Inference}

Once the models are trained, the system is deployed to perform bot detection. Ideally, the system must allow for two modes of execution: (i) model (re)training, to adjust to the dynamics of the network, and (ii) inference, \ie to predict whether or not a given host is a bot. The inference unfolds in two steps---presumable benign hosts get filtered out in Phase 1 as they get assigned to the benign cluster, and suspicious hosts, assigned to a different cluster, are further classified in Phase 2.  
\fig{fig:flowchart} captures the inner workings of node classification. 
To preserve consistency, the deployed system must synchronize and execute requests in order of observation. 

\section{Experiments}\label{sec:experiments}

We evaluate the proposed bot detection system on a Hadoop cluster. In this section, we detail the experimental setup and the results of our evaluation.
\subsection{Environment Setup}
\subsubsection{Hardware}
We experiment on a Hadoop cluster that consists of a management node (2x Intel Xeon Silver 4114; 192 GB RAM), a compute node (2x Intel Xeon Gold 5120; 384 GB RAM) and four data nodes (2x Intel Xeon Silver 4114; 192 GB RAM). 
A 25Gbit and 10Gbit physical networks are deployed, interconnecting the nodes. The former network is primarily used for data and applications, while the latter is for administration.
\vspace{-1mm}
\subsubsection{Software}
The software implementation is primarily based on Java. To ease dependency management, the project incorporates Gradle~\cite{gradle}. JGraphT~\cite{jgrapht} graph library is used to construct the graph and extract graph-based features from network flows. Both Smile~\cite{smile} and Encog~\cite{encog} are used in tandem for ML. In order to support rapid prototyping, a custom in-house DataFrame (DF) library has been developed. DF conforms to the incremental streaming paradigms, data streams with well-defined source, pipelines and sinks. Furthermore, the underlying data structures are immutable and all the basic stream-based transformations are available.

\subsection{Dataset}
Our evaluation is based on the CTU-13~\cite{ctu13} dataset. CTU-13 comprises of 13 different subset datasets (DS) that include captures from 7 distinct malware, performing port scanning, DDoS, click fraud, spamming, \etc
\begin{table}[htbp!]
\centering
\caption{CTU-13 Dataset}
\begin{tabular}{|c|c|c|c|c|} 
\hline
DS & Duration & \# Flows & Bot & \# Bots \\ [0.5ex] 
\hline\hline
1 & 6.15 & 2824637 & Neris & 1\\
\hline
2 & 4.21 & 1808123 & Neris & 1\\
\hline
3 & 66.85 & 4710639 & Rbot & 1\\
\hline
4 & 4.21 & 1121077& Rbot & 1\\
\hline
5 & 11.63 & 129833 & Virut & 1\\
\hline
6 & 2.18 & 558920 & Menti & 1\\
\hline
7 & 0.38 & 114078 & Sogou & 1\\
\hline
8 & 19.5 & 2954231 & Murlo & 1\\
\hline
9 & 5.18 & 2753885 & Neris & 10 \\
\hline
10 & 4.75 & 1309792 & Rbot & 10\\
\hline
11 & 0.26 & 107252 & Rbot & 3\\
\hline
12 & 1.21 & 325472 & NSIS.ay & 3\\
\hline
13 & 16.36 & 1925150 & Virut & 1\\
\hline
\end{tabular}
\label{tab:ctu13_ds}
 \end{table}
Every subset carries a unique network topology with a certain number of bots that leverage different protocols. Table~\ref{tab:ctu13_ds} summarizes the dataset duration, number of flows and bots, and the type of bot in every subset. CTU-13 labels indicate whether a flow is from/to botnet, background or benign. Therefore, known infected hosts are labeled as bots and remaining hosts as benign. We leverage 12 datasets as base training data, while a single dataset, \#9, is left out for testing purpose. This test dataset contains NetFlow data collected from a Neris botnet, 10 unique hosts labeled as bots, performing multiple actions. We use this dataset configuration for training and testing, unless stated otherwise.



\subsection{Results and Discussion}
\subsubsection{Graph Transform, Feature Extraction \& Normalization}
For every subset in the CTU-13 dataset, the system first ingests all the network flows, creates the graph, extracts base features and normalizes them. For each dataset,~\tab{tab:gt_runtime} highlights the graph creation time \ie graph transform (GT), number of graph nodes ($|V|$), total runtime to extract only base BC feature and all base features (FE), and total runtime to normalize features (F-Norm).
\begin{table}[htbp!]
\centering
\scriptsize
\caption{Graph Transform, Base Feature Extraction and Normalization Computation}
\begin{tabular}{|c|c|c|c|c|c|} 
\hline
DS & GT & Nodes & BC & FE & F-Norm\\
& (seconds) & & (hours) & (hours) & (seconds)\\ [0.5ex] 
\hline\hline
1 & 9 & 606829 & 24.12 & 24.121 & 11.3\\
\hline
2 & 6 & 441845 & 10.387 & 10.624 & 7.9\\
\hline
3 & 21 & 434489 & 9.463 & 9.713 & 13.755\\
\hline
4 & 5 & 185742 & 1.37 & 1.431 & 6.307\\
\hline
5 & 1 & 41548 & 0.057 & 0.06 & 0.556\\
\hline
6 & 3 & 107056 & 0.28 & 0.295 & 2.112\\
\hline
7 & 1 & 38081 & 0.021 & 0.022 & 0.488\\
\hline
8 & 13 & 383339 & 9.67 & 9.954 & 9.617\\
\hline
9 & 10 & 366881 & 8.677 & 8.97 & 7.879\\
\hline
10 & 7 & 197542 & 1.06 & 1.108 & 4.861\\
\hline
11 & 1 & 41809 & 0.055 & 0.057 & 0.627\\
\hline
12 & 2 & 94164 & 0.287 & 0.302 & 1.412\\
\hline
13 & 9 & 315343 & 3.667 & 3.852 & 6.824\\
\hline
\end{tabular}
\label{tab:gt_runtime}
 \end{table}

It is evident that there is a non-linear relationship between BC and the number of nodes in the graph. Furthermore, the inconsistent variation between GT and the number of nodes is due to the differing time windows across datasets. Also, dataset \#3 has a much higher number of flows than \#2, which increases the runtime of graph creation. This is primarily due to the repeated modification of exclusive flow tuples in set $A$. 
The system then normalizes the base features, and~\tab{tab:gt_runtime} depicts 
its total runtime with $D=1$. Evidently, normalizing features does not significantly increase the total runtime of the system. The largest runtime reported for the most complex dataset is 13.755 seconds.  

\subsubsection{Stand-alone SL}
We start by highlighting the limitations of a stand-alone supervised learning approach. This consists of evaluating supervised ML classifiers, including DT, LR, SVM and FNN for bot detection. Each classifier employs graph-based normalized features and is trained on the entire training dataset. In our experiments, DT uses the Gini instance split rule algorithm, LR is used without regularization, and SVM uses the Gaussian kernel with a soft margin penalty of 1. Moreover, NN is configured to use cross entropy as an error function and 10 hidden layers of 1000 units each.
Table \ref{tab:sl_singlephase} highlights the results, where LR and DT show meaningful classification. Both LR and DT classifiers result in a 100\% recall, with 91\% and 83\% precision, respectively. With LR's superiority in precision, it \textit{seems to be} the classifier of choice. 
The other classifiers were able to accurately classify all the benign hosts, but failed to identify any bots.

\begin{table}[htbp!]
\centering
\caption{Supervised Learning with F-Norm}
\begin{tabular}{|c|c|c|c|c|c|c|} 
\hline
Classifier & TP & FP & TN & FN & Recall & Precision\\[0.5ex] 
\hline\hline
DT & 10 & 2	& 366869 & 0 & 100 & 83\\
\hline
\rowcolor{lightgray}
LR & 10 & 1 & 366870 & 0 & 100 & 91\\
\hline
SVM & 0 & 0 & 366871 & 10 & 0 & 0\\
\hline
FNN & 0 & 0 & 366871 & 10 & 0 & 0\\
\hline
\end{tabular}
\label{tab:sl_singlephase}
 \end{table}

We then evaluate the training time and robustness of the stand-alone classifiers, as depicted in Tables \ref{tab:sl_trainingtime} and \ref{tab:sl_singlephase_robustness}. 
DT requires the least training time of 4.9 seconds, which is in high contrast to the 58.2 seconds for LR. 
That is, DT requires only 8.4\% of LR's training time for the entire training dataset. It is also essential for a bot detection system to detect bots that the classifier has never seen before \ie unknown or zero-day attacks. 
Therefore, to evaluate robustness to zero-day attacks, we change the selection of the training and testing datasets. We choose dataset \#6 for testing, which has a unique bot that is not present in any other dataset. The remaining datasets are aggregated to form the training set, with 34 bots and a total of $\approx$3.1M hosts. 
Evidently, DT outperforms LR, which misclassifies a benign host, with a low precision of 50\%.

\begin{table}[htbp!]
\centering
\caption{Training Time of Supervised ML Classifiers}
\begin{tabular}{|c|c|} 
\hline
Classifier & Training Time (s)\\[0.5ex] 
\hline\hline
\rowcolor{yellow}
DT & 4.9\\
\hline
LR & 58.2\\
\hline
SVM & 6832.3\\
\hline
FNN & 5.7\\
\hline
\end{tabular}
\label{tab:sl_trainingtime}
\end{table}
\begin{table}[htbp!]
\centering
\caption{Supervised Learning Against Previously Unknown Bot}
\begin{tabular}{|c|c|c|c|c|c|c|} 
\hline
Classifier & TP & FP & TN & FN & Recall & Precision\\[0.5ex] 
\hline\hline
\rowcolor{yellow}
DT & 1 & 0 & 107055 & 0 & 100 & 100\\
\hline
\rowcolor{lightgray}
LR & 1 & 1 & 107054 & 0 & 100 & 50\\
\hline
SVM & 0 & 0 & 107055 & 1 & 0 & 0\\
\hline
FNN & 0 & 0 & 107055 & 1 & 0 & 0\\
\hline
\end{tabular}
\label{tab:sl_singlephase_robustness}
\end{table}
\vspace{3mm}
Based on the above evaluations, LR outperforms DT in precision, while DT shows a superior training time and robustness to unknown attacks. However, precision, training time and robustness are all crucial for our bot detection system. Can we achieve the best of all three? To investigate this, we set out to evaluate a two-phased system that employs an initial clustering phase (UL), followed by a classification phase (SL). We delineate its evaluation in the following subsections.

\subsubsection{Phase 1 (UL)}
For this phase of the system, we evaluate three UL techniques, namely $k$-Means, DBScan and SOM. However, DBScan results are inconclusive, where bots co-located with benign hosts. In essence, DBScan does not produce a single, prevalent benign cluster. DBScan is evaluated with varying minimum number of neighborhood points (minPts) and distance ($\epsilon$). Multiple $\epsilon$ values are tested in the range of [10\textsuperscript{-5}, 10\textsuperscript{-4}, ..., 10\textsuperscript{5}]. Also, we infer $\epsilon$ values that correspond to the boundary of the bots themselves. We vary minPts in [1, 2, ..., 25] depending on the number of bots in the aggregated training dataset. However, maximal separation of bots from benign hosts could not be achieved with the tested parameters.

On the other hand, both $k$-Means and SOM show appreciable results, where SOM is trained with a learning rate of 0.7. 
Tables \ref{tab:kmeans} and \ref{tab:som1} highlight the evaluation metrics, including number of clusters or neurons, number of hosts outside the benign cluster (HOB), percentage of hosts outside the benign cluster relative to the total number of hosts (HOB\%), number of bots outside the benign cluster (BOB), and percentage of bots relative to the total number of bots (BOB\%).

Recall, the dataset \#9 is removed for testing, which includes 10 hosts labeled as bots and $\approx$366K hosts. Also, $\approx$3.2M hosts from the remaining datasets are used to train the classifiers. 
In comparison to the number of clusters for $k$-Means, SOM is able to alienate its first bot outside the benign cluster with a lower number of neurons (9 \textit{vs.} 16). 
With 81 neurons, SOM has a recall rate of 92\%, with $k$-Means achieving 42\%. However, $k$-Means catches up with 121 clusters. Nevertheless, SOM outperforms $k$-Means by maximizing the number of bots isolated with a smaller number of neurons.

\begin{table}[htbp!]
\centering
\caption{$k$-Means Clustering with F-Norm}
\begin{tabular}{|c|c|c|c|c|} 
\hline
\# of Clusters & HOB & HOB\% & BOB & BOB\% \\ [0.5ex] 
\hline\hline
4 & 5 & 0.0002 & 0 & 0\\
\hline
9 & 12 & 0.0004 & 0 & 0\\
\hline
16 & 36 & 0.0012 & 1 & 4\\
\hline
25 & 94 & 0.0033 &  6 & 24\\
\hline
36 & 170 & 0.0059 &  6 & 24 \\
\hline
49 & 473 & 0.0164 &  8 & 32\\
\hline
64 & 1071 & 0.0371 & 10 & 40\\
\hline
81 & 1133 & 0.0392 & 10 & 40\\
\hline
\rowcolor{yellow}
100 & 3028 & 0.1049 & 21 & 87.5\\
\hline
121 & 26935 & 0.9327 & 24 & 96\\
\hline
144 & 27100 & 0.9384 & 24 & 96\\
\hline
169 & 27302 & 0.9454 & 24 & 96\\
\hline
196 & 27359 & 0.9474 & 24 & 96\\
\hline
225 & 28752 & 0.9956 & 24 & 96\\
\hline
\end{tabular}
\label{tab:kmeans}
\end{table}

\begin{table}[htbp!]
\centering
\caption{SOM Clustering with F-Norm}
\begin{tabular}{|c|c|c|c|c|} 
\hline
\# of Neurons & HOB & HOB\% & BOB & BOB\% \\ [0.5ex] 
\hline\hline
4 & 10 & 0.0004 & 0 & 0\\
\hline
9 & 29 & 0.0010 & 1 & 4\\
\hline
16 & 49 & 0.0017 & 1 & 4\\
\hline
25 & 113 & 0.0039 & 6 & 24\\
\hline
36 & 286 & 0.0099 & 7 & 28 \\
\hline
49 & 556 & 0.0193 & 8 & 32\\
\hline
64 & 1709 & 0.0592 & 10 & 40\\
\hline
81 & 3524 & 0.1222 & 23 & 92\\
\hline
\rowcolor{yellow}
100 & 3675 & 0.1274 & 23 & 92\\
\hline
121 & 3894 & 0.1350 & 23 & 92\\
\hline
144 & 27591 & 0.9647 & 24 & 96\\
\hline
169 & 27856 & 0.9740 & 24 & 96\\
\hline
196 & 28342 & 0.9912 & 24 & 96\\
\hline
225 & 28449 & 0.9950 & 24 & 96\\
\hline
\end{tabular}
\label{tab:som1}
 \end{table}
 
\vspace{2mm}
With a cluster size of 100, $k$-Means alienates 21 bots, while having an outside host sum of 3028 for the remaining non-benign clusters. In contrast, SOM removes 23 bots from the benign cluster with an outside host sum of 3675. The very next $k$-Means cluster size \ie 121, boosts HOB from 3028 to 26935, while SOM remains at a close 3894. However, $k$-Means isolates three extra bots, yielding 24 BOB for 26935 HOB. That is, three extra bots were detected for a $\approx$23K increase in HOB. 
Recall, our objective in this phase is to jointly minimize HOB while maximizing BOB. Therefore, SOM with 100 neurons becomes the natural choice.

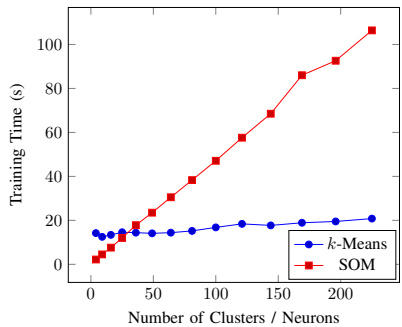
\begin{figure}
\centering
\resizebox{0.29\textwidth}{!}{%
  \begin{tikzpicture}
\begin{axis}[
	xlabel=Number of Clusters / Neurons,
	ylabel=Training Time (s),
    legend style={at={(0.99,0.1)},anchor=east}
]
\addplot 
	coordinates {
    (4,14.2)
(9,12.5)
(16,13.4)
(25,14.6)
(36,14.4)
(49,14.1)
(64,14.4)
(81,15.2)
(100,16.8)
(121,18.4)
(144,17.7)
(169,18.9)
(196,19.5)
(225,20.8)

};
\addplot 
	coordinates {
    (4,2.2)
(9,4.5)
(16,7.6)
(25,12)
(36,17.8)
(49,23.5)
(64,30.5)
(81,38.3)
(100,47.1)
(121,57.6)
(144,68.5)
(169,86)
(196,92.6)
(225,106.4)

};
\legend{$k$-Means,SOM}
\end{axis}
\end{tikzpicture}}
  \caption{Comparison of SOM and $k$-Means with respect to training time}
  \label{fig:runtime_km_som}
\end{figure}

With respect to runtime, $k$-Means mostly outperforms SOM, as depicted in~\fig{fig:runtime_km_som}. With 100 clusters, $k$-Means took 16.8 seconds to train, in comparison to 47.1 seconds of SOM.  
We speculate that SOM's ever increasing training time is contributed to how 
it updates the surrounding neurons. As the number of neurons increases, the density of their neighborhood increases. Eventually, more neurons will tend to be within the threshold radius. Nevertheless, with recall being our top priority, we leverage SOM as the UL classifier in Phase 1.

\vspace{3mm}
\subsubsection{Phase 2 (SL)}

The training set for Phase 2 is determined by the number of hosts outside the benign cluster in Phase 1. 
These are the relevant hosts for this phase, as hosts that are assigned in the benign cluster never make it to Phase 2. With a 10$\times$10 (\ie 100 neurons) SOM and normalized features in Phase 1, the size of the dataset is significantly reduced. Therefore, we have 3675 HOB, including 23 bots, for classification in Phase 2. 

\begin{table}[htbp!]
\centering
\caption{Supervised Learning with F-Norm}
\begin{tabular}{|c|c|c|c|c|c|c|} 
\hline
Classifier & TP & FP & TN & FN & Recall & Precision\\ [0.5ex] 
\hline\hline
\rowcolor{yellow}
DT & 10 & 1 & 366870 & 0 & 100 & 90.9\\
\hline
LR & 0 & 0 & 366871 & 10 & 0 & 0\\
\hline
SVM & 0 & 0 & 366871 & 10 & 0 & 0\\
\hline
FNN & 0 & 0 & 366871 & 10 & 0 & 0\\
\hline
\end{tabular}
\label{tab:sp_results}
\end{table}
\vspace{3mm}
For this phase of the system, we evaluate four SL techniques, namely DT, LR, SVM and FNN.  
The DT classifier shows the best performance with the small dataset, as depicted in \tab{tab:sp_results}. It successfully detects \textit{all} bots in the test dataset, with only a single FP out of the 366871 benign hosts. In contrast, all other classifiers are lackluster and unable to recall even a single bot from the dataset. We believe this is because all classifiers, except DT, rely on gradient-descent for error-correction. This implies that every single node in the dataset will affect the end-hypothesis function.  
Thus, with a dataset that is unbalanced, the hypothesis will be biased towards the benign hosts, which is the case for LR, SVM and FNN.

Table \ref{tab:hybrid_sl_trainingtime} highlights the training time for the supervised  classifiers. 
For Phase 1, a 10$\times$10 SOM incurs a training time of 
47.1 seconds, while DT has the lowest training time of 88 milliseconds in Phase 2. Thus, the aggregate training time for both phases is $\approx$47.2 seconds. This is an 11 seconds improvement over the 58.2 seconds previously observed for a stand-alone LR classifier.


\begin{table}[htbp!]
\centering
\caption{Training Time of Supervised Classifiers on the Pruned Dataset}
\begin{tabular}{|c|c|} 
\hline
Classifier & Training Time (ms)\\[0.5ex] 
\hline\hline
\rowcolor{yellow}
DT & 88\\
\hline
LR & 2454\\
\hline
SVM & 864\\
\hline
NN & 22\\
\hline
\end{tabular}
\label{tab:hybrid_sl_trainingtime}
 \end{table}

\vspace{3mm}
Using dataset \#6 for testing, the robustness test harbors more hosts for training in Phase 2. Most importantly, there are more BOBs (\ie 32) and relatively the same HOBs, yielding a higher ratio of bots to hosts outside the benign cluster. The robustness results are portrayed in Table \ref{tab:sl_singlephase_robustness}. Though LR is able to recall the malicious bot while incurring only a single FP, DT exhibits perfect results on this specific test dataset. It is able to detect the previously unknown bot, as well as correctly classify all the benign hosts. Therefore, with SOM selected for Phase 1 and DT for Phase 2, the system ensures minimal training time and robustness to unknown attacks, with high recall and precision.

\subsubsection{Feature Normalization}
Recall that aggregating datasets from different networks can negatively impact the base features, thus compromising system performance. 
Essentially, the topological structure of different networks affect the extracted graphical features, greatly skewing bot pattern and behavior. Thus, the intuition behind feature normalization is to make hosts, including bots, from different datasets look alike.

\tab{tab:som_wo_nfr} showcase the crucial depreciation of the SOM results without normalizing graph-based features. For example, with 81 neurons, SOM with and without F-Norm scores 92\% and 60\% on BOB, respectively. On average, the results without F-Norm have a higher HOB. This intrinsic observation signifies the lack of similarity between hosts of the same category. For example, benign hosts from different networks are not co-located due to the stark differences in their features. Conversely, with F-Norm, similarly labeled hosts are more frequently co-located, yielding better BOB and HOB. Hence, normalized graph-based features significantly improve the spatial stability of ML in the bot detection system.

\begin{table}[htbp!]
\centering
\caption{SOM Clustering without F-Norm}
\begin{tabular}{|c|c|c|c|c|} 
\hline
\# of Neurons & HOB & HOB\% & BOB & BOB\% \\ [0.5ex] 
\hline\hline
4 & 8 & 0.0003 & 0 & 0\\
\hline
9 & 39 & 0.0014 & 0 & 0\\
\hline
16 & 689 & 0.0239 & 0 & 0\\
\hline
25 & 935 & 0.324 & 0 & 0\\
\hline
36 & 2280 & 0.0790 & 9 & 36\\
\hline
49 & 3792 & 0.1315 & 11 & 44\\
\hline
64 & 4207 & 0.1459 & 14 & 56\\
\hline
81 & 6721 & 0.2333 & 15 & 60\\
\hline
\rowcolor{yellow}
100 & 8465 & 0.2940 & 22 & 88\\
\hline
121 & 12923 & 0.4495 & 24 & 96\\
\hline
144 & 20780 & 0.7248 & 24 & 96\\
\hline
169 & 22607 & 0.7890 & 24 & 96\\
\hline
196 & 23714 & 0.8280 & 24 & 96\\
\hline
225 & 42125 & 1.4803 & 24 & 96\\
\hline
\end{tabular}
\label{tab:som_wo_nfr} \end{table}

\begin{figure}
\centering
\resizebox{0.29\textwidth}{!}{%
\begin{tikzpicture}
\begin{axis}[
	xlabel=Number of Neurons,
	ylabel=BOB,
    legend style={at={(0.99,0.1)},anchor=east}
]
\addplot 
	coordinates {
(4,0)
(9,1)
(16,1)
(25,6)
(36,7)
(49,8)
(64,10)
(81,23)
(100,23)
(121,23)
(144,24)
(169,24)
(196,24)
(225,24)

};
\addplot 
	coordinates {
    (4,0)
(9,0)
(16,0)
(25,0)
(36,9)
(49,11)
(64,14)
(81,15)
(100,22)
(121,24)
(144,24)
(169,24)
(196,24)
(225,24)
};
\draw[fill=yellow] (100, 22.5) circle[radius=7pt];
\legend{With F-Norm, Without F-Norm}
\end{axis}
\end{tikzpicture}}
  \caption{Number of bots outside the benign cluster (BOB) assigned by SOM with and without feature normalization}
  \label{fig:nfr_hob}
\end{figure}
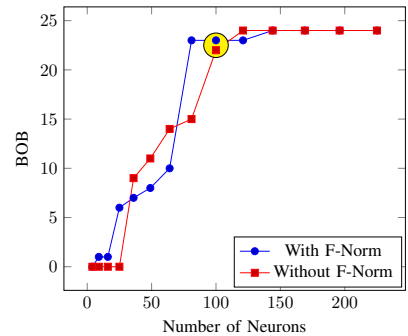

For 100 neurons, SOM with F-Norm results in 23 BOB and 3675 HOB. Without F-Norm, it results in 22 BOB and 8465 HOB, as shown in Figures~\ref{fig:nfr_bob} and~\ref{fig:nfr_hob}. Thus, for the same number of neurons, feature normalization was able to maximize BOB, while minimizing HOB. Therefore, we choose 100 neurons with F-Norm as our primary configuration for SOM.

\subsubsection{Feature Engineering}

It is important to gauge the significance and impact of the graph-based features on the hybrid bot detection system. Albeit different feature combinations may impact results, are all features necessary? 
\tab{tab:feat_corr} shows the Pearson's feature correlation matrix for the normalized graph-based features. 
At a glance, we can determine that the first five features are highly correlated, with a correlation close to or greater than 0.9. Therefore, feature combinations that exclude some of these features may not exacerbate classification accuracy. On the other hand, the last two features are highly uncorrelated, with LCC being the least correlated. 
Hence, we start with removing IDW and ODW, which slightly decreases the benign cluster size by $\approx$24K hosts but adds 1 more BOB. However, the performance of the SL classifiers suffer when we eliminate IDW and ODW features. Precision drops to 52.6\% for DT from 90.9\% (\cf \tab{tab:sp_results}). Also, LR now misclassifies two benign hosts as bots.

\begin{figure}
\centering
\resizebox{0.29\textwidth}{!}{%
  \begin{tikzpicture}
\begin{axis}[
	xlabel=Number of Neurons,
	ylabel=HOB,
    legend style={at={(0.99,0.1)},anchor=east}
]
\addplot 
	coordinates {
(4,10)
(9,29)
(16,49)
(25,113)
(36,286)
(49,556)
(64,1709)
(81,3524)
(100,3675)
(121,3894)
(144,27591)
(169,27856)
(196,28342)
(225,28449)
};
\addplot 
	coordinates {(4,8)
(9,39)
(16,689)
(25,935)
(36,2280)
(49,3792)
(64,4207)
(81,6721)
(100,8465)
(121,12923)
(144,20780)
(169,22607)
(196,23714)
(225,42125)
};
\draw[fill=yellow] (100, 8465) circle[radius=5pt];
\draw[fill=yellow] (100, 3675) circle[radius=5pt];
\legend{With F-Norm,Without F-Norm}
\end{axis}
\end{tikzpicture}}
  \caption{Number of hosts outside the benign cluster (HOB) assigned by SOM with and without feature normalization}
  \label{fig:nfr_bob}
\end{figure}
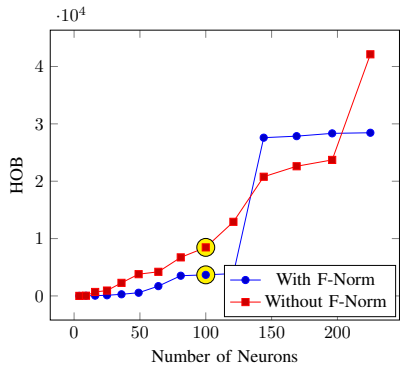

\begin{table}[htbp!]
\centering
\caption{Pearson's Feature Correlation Matrix with F-Norm}
\begin{tabular}{|l|c|c|c|c|c|c|c|} 
\hline
& ID & IDW & OD & ODW & BC & LCC & AC\\ [0.5ex] 
\hline\hline
ID & 1 & \cellcolor{yellow}0.99 & \cellcolor{yellow}0.92 & \cellcolor{yellow}0.95 & \cellcolor{yellow}0.96 & 0.03 & 0.32\\
\hline
IDW & 0.99 & 1 & \cellcolor{yellow}0.91 & \cellcolor{yellow}0.96 & \cellcolor{yellow}0.97 & 0.03 & 0.33\\
\hline
OD & 0.92 & 0.91 & 1 & \cellcolor{yellow}0.89 & \cellcolor{yellow}0.90 & 0.08 & 0.37\\
\hline
ODW & 0.95 & 0.96 & 0.89 & 1 & \cellcolor{yellow}0.97 & 0.04 & 0.43\\
\hline
BC & 0.96 & 0.97 & 0.90 & 0.97 & 1 & 0.01 & 0.46 \\
\hline
LCC & 0.03 & 0.03 & 0.08 & 0.04 & 0.01 & 1 & 0.01\\
\hline
AC & 0.32 & 0.33 & 0.37 & 0.43 & 0.46 & 0.01 & 1\\
\hline
\end{tabular}
\label{tab:feat_corr}
\end{table}
\hphantom\\\hphantom\\
A weakness of the chosen features is the runtime of BC. For the first dataset, it took over 24 hours to compute BC. This will impede any effort to expedite the learning process. Without BC, SOM maintains similar performance. However, removing BC from the feature set adversely affects the precision of DT, dropping to 62.5\%. SOM without BC performs identical to the use of the entire feature set. In contrast, DT's precision is affected by the removal of BC, but it is better than that of the removal of IDW and ODW. While the precision deteriorated, \textit{only} 6 and 9 benign hosts were misclassified out of the $\approx$367K hosts with the removal of BC and IDW/ODW, respectively. This reinforces the correlation matrix \ie having these features the most correlated. Since recall and precision are sought after metrics in our system, it is important to include these features for training and testing classifiers.
\vspace{-2mm}
\section{Conclusion}\label{sec:conclusion}
The struggle to detect malicious agents in a network has recently converged to ML. High FPs and FNs are detrimental to any intrusion detection system. 
Network-based approaches exhibit plausible detection rates. When paired with a proper modeling technique, such as graphs, high detection accuracy can be achieved with low FPs. In this paper, we propose a two-phased, graph-based bot detection system that leverages both supervised and unsupervised learning.

Using SOM, Phase 1 establishes a compromise between maximizing the benign cluster and alienating the malicious bots. Furthermore, the results of Phase 2 favor DT, showcasing high TPs and low FPs. Moreover, feature normalization significantly improves the spatial stability of the models. 
The system is robust against unknown attacks and cross-network ML model training and inference. It detects bots that rely on different protocols and is suitable for large-scale data.


\section*{Acknowledgments}
This work is supported in part by the Royal Bank of Canada, and the NSERC CRD Grant No. 530335.
\newpage
\bibliographystyle{IEEEtran} 

\end{document}